# SHORT SPACING SYNTHESIS FROM A PRIMARY BEAM SCANNED INTERFEROMETER

**R.D. Ekers[1] and A.H. Rots[2]**

[1]Kapteyn Astronomical Institute, University of Groningen,
Groningen, the Netherlands.

[2]Netherlands Foundation for Radio Astronomy, Dwingeloo,
the Netherlands.

## 1. INTRODUCTION

Aperture synthesis instruments providing a generally highly uniform sampling of the visibility function often leave an unsampled hole near the origin of the $(u,v)$-plane. One of the common configurations of the Westerbork Synthesis Radio Telescope (WSRT) provides sampling in concentric rings with radii of $n \times 18$ m. However, the samples with $n = 0$ and $n = 1$ cannot be obtained directly with the instrument because of rather obvious physical constraints. This missing short spacing information causes a central depression of the synthesized beam pattern, resulting in large negative, bowl-shaped areas around (especially) extended sources. Very large sources will only partially be present in the maps. Annoying as this can be for continuum observations, in the case of spectral line data it causes serious distortions of the individual line profiles.

In this paper we shall use the following nomenclature. Coordinates on the sky are $l$ and $m$; in the aperture plane $u$ and $v$. Capital letters will indicate functions in the plane of the sky, lower case letters their Fourier Transformations in the $(u,v)$-plane. $B$ is the brightness distribution on the sky, $A$ the antenna pattern of the individual 25 m elements of the array (the so-called primary beam), and $S$ is a bed-of-nails sampling function on the sky; $\delta$ signifies the usual delta function. Multiplications are represented by a dot ($\cdot$), convolutions by an asterisk ($*$).

If $t(u,v)$ is the sampling of the entire array, then we have observed:

$$b'(u,v) = \bigl(b(u,v) * a(u,v)\bigr) \cdot t(u,v) \tag{1$^\text{a}$}$$

resulting in a map:

$$B'(l,m) = \bigl(B(l,m) \cdot A(l,m)\bigr) * T(l,m) \tag{1$^\text{b}$}$$

61

The information we would like to add to the map is:

$$b'(0,0) = \bigl(b(u,v) * a(u,v)\bigr) \cdot \delta(u,v) = \iint B(l,m) \cdot A(l,m)\, dl\, dm \quad (2)$$

for the missing zero spacing flux, and:

$$b'(18\text{m}) = \bigl(b(u,v) * a(u,v)\bigr) \cdot t_{18}(u,v) \quad (3)$$

where $t_{18}$ represents the sampling along a hypothetical 18 m spacing track.

## 2. A SINGLE TELESCOPE

Just one observation with one telescope, at the proper position, will only provide the zero spacing flux as expressed in Eq. 2. It has been recognized, however, for many years that one may obtain more information about the visibility function by scanning the whole map area in the sky with a single telescope. In that way one can theoretically derive the visibility function in a circle with a radius equal to the diameter of the dish, provided one knows the antenna pattern $A_s$ of this telescope. The result of the scanning is:

$$B''(l,m) = (B(l,m) * A_s(l,m)) \cdot S_s(l,m) \quad (4^a)$$

which transforms into (neglecting $S_s$):

$$b''_s(u,v) = b(u,v) \cdot a_s(u,v) \quad (4^b)$$

It is quite easy to see that one can obtain the desired functions in Eq. 2 and 3 from these data.

Usually one requires a dish with a diameter of at least 60 m to fill in the hole in WSRT observations.

## 3. SHORT BASELINE INTERFEROMETER

Single observations with a short baseline interferometer (e.g., 36 m) only give visibility data along the track of the interferometer in the $(u,v)$-plane. In analogy with the single dish case it is possible, however, to get information about the visibility function in a broad annulus around that track by scanning an area in the sky.

Assuming that the present orientation of the interferometer corresponds to $u = u_0$, $v = v_0$; leaving the delay center at $l = 0$, $m = 0$; and pointing the telescopes into the direction $l = l'$, $m = m'$, the output of the interferometer is:

$$\iint B(l,m) \cdot A(l-l', m-m') \cdot e^{2\pi i(u_0 l + v_0 m)} \, dl \, dm =$$
$$= \{b(u,v) * (a(u,v) \cdot e^{-2\pi i(ul' + vm')})\} \cdot \delta(u-u_0, v-v_0) =$$
$$= \iint b(u_0 - u, v_0 - v) \cdot a(u,v) \cdot e^{-2\pi i(ul' + vm')} \, du \, dv =$$
$$= \{(B(-l,-m) \cdot e^{-2\pi i(u_0 l + v_0 m)}) * A(l,m)\} \cdot \delta(l-l', m-m') \quad (5)$$

If one observes this quantity on a grid $S(l', m')$ and performs the Fourier Transformation over $l'$ and $m'$, one obtains:

$$b'''(u,v) = b(u_0 - u, v_0 - v) \cdot a(u,v) \quad (6)$$

## 4. RETRIEVING SHORT SPACING INFORMATION

Scanning the sky with the 36 m baseline interferometer and recording the cross correlation as well as the total power of (one of) the telescopes yields $b''_s = b''$ (since $a_s = a$) from Eq. 4[b] and $b'''$ from Eq. 6. The telescopes are 25 m in diameter and hence $b''$ extends to a radius of 25 m, while $b'''$ is defined in an annulus between radii 36±25 m. Although in practice the overlap between $b''$ and $b'''$ will be smaller, it should nevertheless be possible to reconstruct $b(u,v)$ in the area of interest such that $b'(0)$ and $b'(18m)$ can be constructed.

If one may assume that

$$a(u,v) = p(u) \cdot p(v) \quad (7)$$

– as would be the case if we can approximate $a(u,v)$ by a circular Gaussian –, one only needs scanning in one dimension. Let us for simplicity assume that $v_0 = 0$. Then $m' = 0$ and one has to scan in $l$-direction. Fourier Transformation over $l'$ yields (instead of Eq. 6):

$$b''''(u) = p(u) \cdot \int b(u_0 - u, v) \cdot p(v) \, dv \quad (8)$$

which already contains half of the convolution needed in Eq. 3.